\documentclass[12pt]{article}
\def\linecite#1{[\,\lower0.5em\hbox{\Large\protect\cite{#1}}]}

\def\ra{\rightarrow}

\def\be{\begin{equation}}
\def\beq{\begin{equation}}
\def\ee{\end{equation}}
\def\eeq{\end{equation}}
\def\bea{\begin{eqnarray}}
\def\eea{\end{eqnarray}}
\def\Jpsi{J\kern-0.12em/\kern-0.16em\psi}
\def\mythinspace{\kern0.05em}
\usepackage[figuresleft]{rotating}

\begin{document}
\begin{titlepage}
\thispagestyle{empty}
\begin{center}
\strut
\\
{\Large {\bf Doubly Heavy Tetraquarks and Baryons}\footnote{Invited
plenary lecture at 
International Conference on New Frontiers in Physics, Aug. 28--Sep. 2013, 
Kolymbari, Crete, Greece.}}
\\
\strut
\\
\strut
\\
{\bf Marek Karliner\footnote{email: {\tt marek@proton.tau.ac.il}}}
\\
\strut
\\
{Raymond and Beverly Sackler School of Physics and Astronomy\\
Tel Aviv University,
Tel Aviv,
Israel}
\\
\strut
\\
\strut
\\
\strut
\\
{\bf Abstract}\\
\strut
\\
\end{center}
During the last three years strong experimental evidence from $B$ and charm
factories has been accumulating for the existence of exotic hadronic
quarkonia, narrow resonances which cannot be made from a quark and an
antiquark. Their masses and decay modes show that they contain a heavy
quark-antiquark pair, but their quantum numbers are such that they
must also contain a light quark-antiquark pair. The theoretical
challenge has been to determine the nature of these resonances. The
main possibilities are that they are either ``genuine tetraquarks",
i.e. two quarks and two antiquarks within one confinement volume, or
``hadronic molecules" of two heavy-light mesons.  In the last few months
there as been more and more evidence in favor of the latter. I discuss
the experimental data and its interpretation and provide fairly
precise predictions for masses and quantum numbers of the additional
exotic states which are naturally expected in the molecular picture but
have yet to be observed.  In addition, I provide arguments in favor of the
existence of an even more exotic state -- a hypothetical deuteron-like
bound state of two heavy baryons. 
I also consider  ``baryon-like" states $Q Q' \bar q\bar q'$, 
which if found will be direct evidence not just for near-threshold
binding of two heavy mesons, but for genuine tetraquarks with novel color
networks. I stress the importance of experimental search for doubly-heavy
baryons in this context.
\vfill
\end{titlepage}

\section{First observation of manifestly exotic hadrons}

In late 2007 the Belle Collaboration reported~\cite{Abe:2007tk}
anomalously large partial widths
of $\Upsilon(5S)\ra\Upsilon(2S)$
and $\Upsilon(5S)\ra\Upsilon(1S)$, two orders of magnitude larger
than the analogous decays of $\Upsilon(3S)$.
Soon afterward Harry Lipkin and I proposed~\cite{Karliner:2008rc}
that a four-quark exotic 
resonance $[b\bar b u \bar d]$ 
might mediate these
decays through the cascade
$\Upsilon(mS) \to [b\bar b u \bar d]
 \pi^-\to \Upsilon(nS)\, \pi^+ \pi^-$.
We suggested looking  for the $[b \bar b u \bar d]$
resonance in these decays as peaks in the invariant mass of
$\Upsilon(1S)\,\pi$ or $\Upsilon(2S)\,\pi$ systems.
\begin{figure}[t]
\centerline{\includegraphics[width=0.9\linewidth,clip]{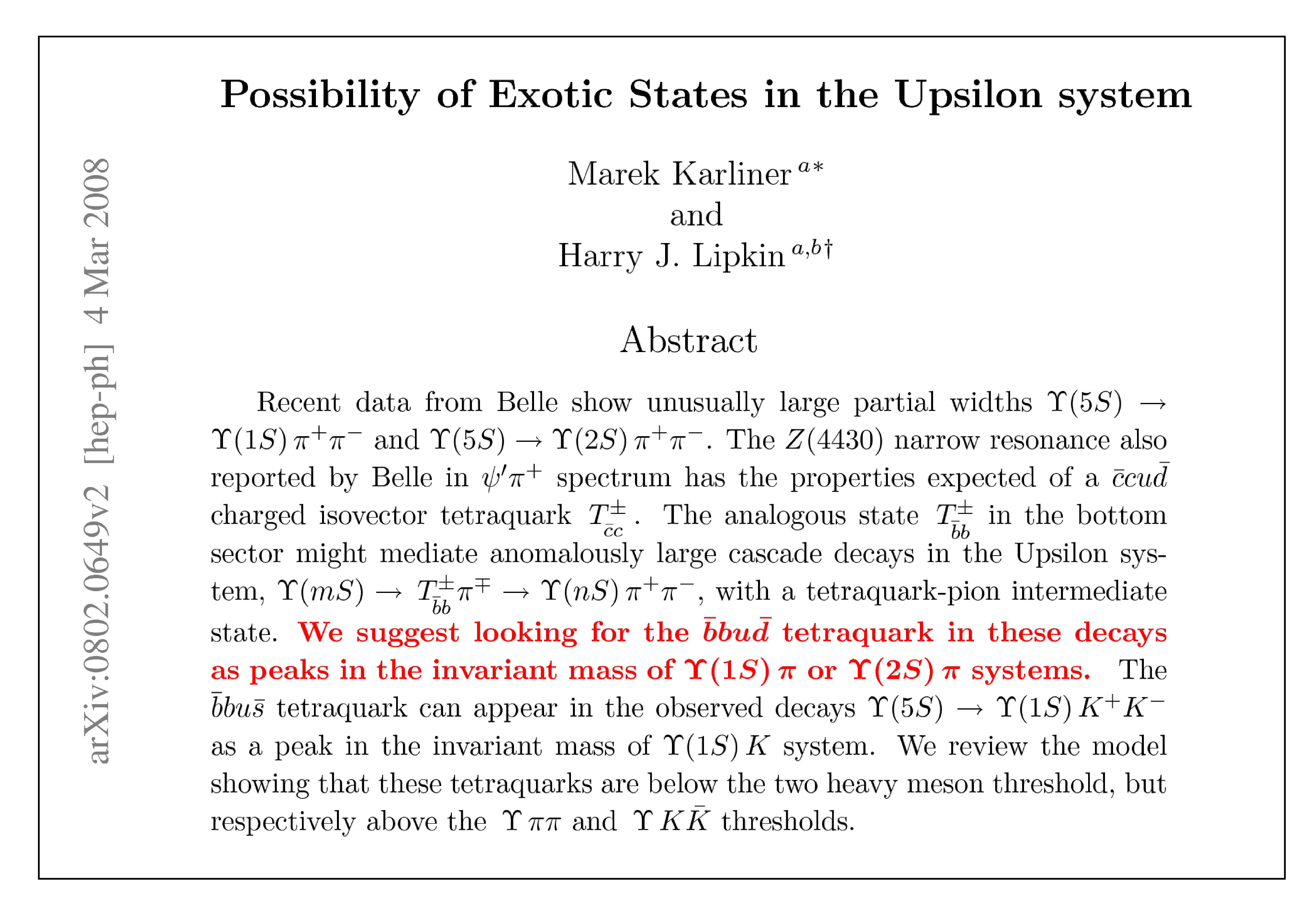}}
\caption{First page of our paper
\protect\cite{Karliner:2008rc}
proposing a four-quark exotic intermediate 
state $b\bar b u \bar d$
as an explanation of the anomalously large rate partial widths
of $\Upsilon(5S)\ra\Upsilon(2S)$
and $\Upsilon(5S)\ra\Upsilon(1S)$.}
\label{fig:KL2008}
\end{figure}
%
More recently Belle collaboration confirmed this prediction, announcing
\cite{Collaboration:2011gja,Belle:2011aa}
the observation
of two charged bottomonium-like resonances $Z_b$ as narrow structures in
\hbox{$\pi^{\pm}\Upsilon(nS)$} \hbox{$(n=1,2,3)$} and 
\hbox{$\pi^{\pm}h_b(mP)$} 
\hbox{$(m=1,2)$} ($h_b$-s are spin-singlet, $P$-wave bottomonia with $J=1$) 
mass spectra that are
produced in association with a single charged pion in $\Upsilon(5S)$
decays.

Since these states decay into bottomonium and a charged pion, 
they must contain both a $\bar b b$ heavy quark-antiquark pair {\em and}
a $\bar d u$ light quark-antiquark pair. Thus their minimal quark content is 
$\bar b b \bar d u$. They are {\em manifestly exotic}.
Their discovery by Belle was the first time such manifestly exotic hadron
resonances have been unambiguously observed experimentally.

The measured masses of the two structures averaged over the five final
states are
\break
$M_1=10608.4\pm2.0$ MeV,
\,$M_2=10653.2\pm1.5$ MeV, both with a width of about 15 MeV.
Interestingly enough,
the two masses $M_1$ and $M_2$ are about 3 MeV above the
respective $B^* \bar B$ and $B^* \bar{B^*}$ thresholds.

\begin{figure}[t]
\centerline{
\fbox{
{\includegraphics[width=0.3\linewidth,clip]{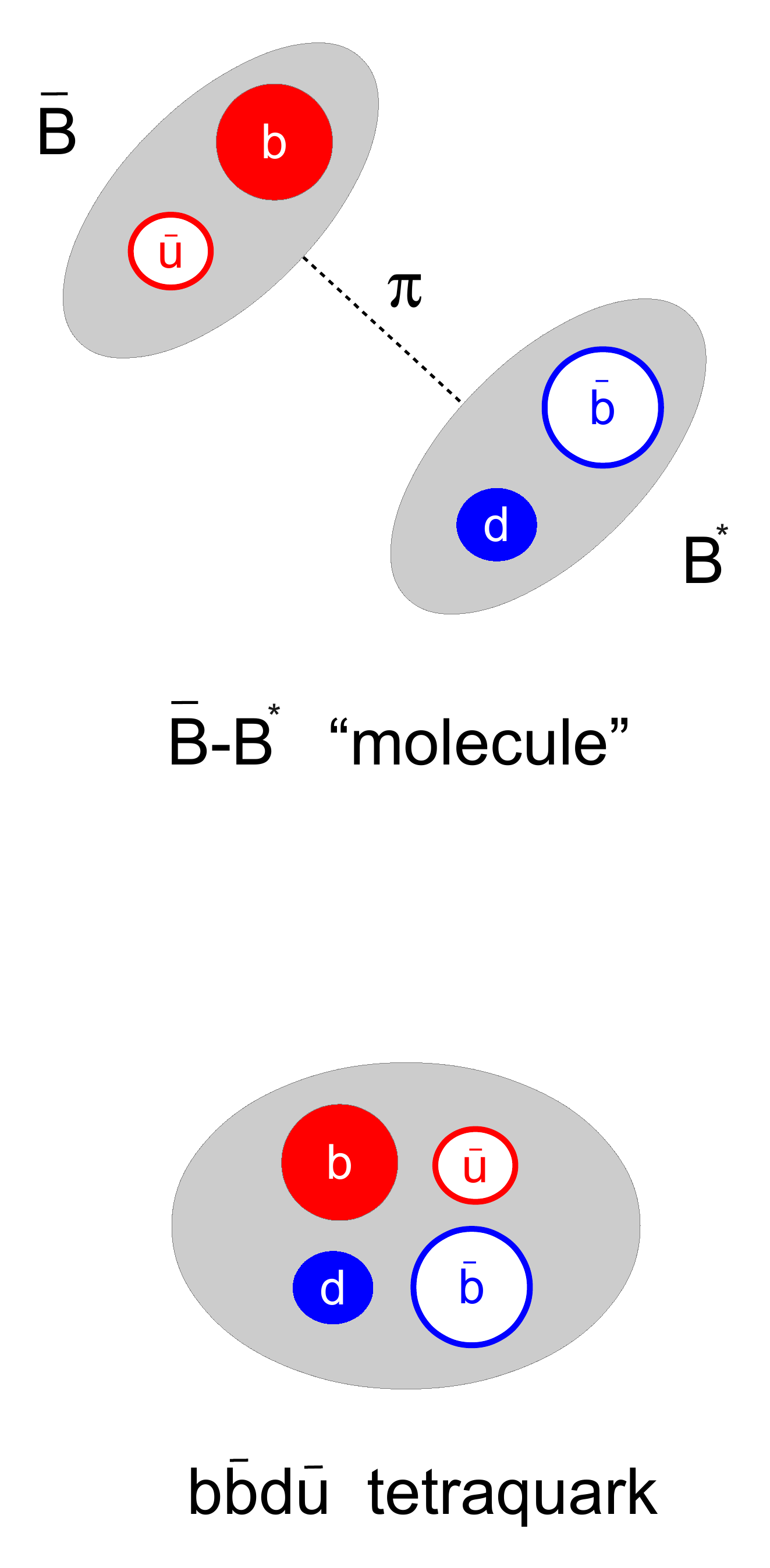}}
}}
\strut
\caption{A schematic depiction of a $\bar B$--$B^*$ deuteron-like
``hadronic molecule" vs. a $b\bar b d \bar u$ tetraquark.
Note that, unlike in the ``molecule",
 in the tetraquark configuration the $b\bar u$ and $d\bar b$
diquarks in general won't be color singlets.}
\label{fig:tetraquark_vs_molecule}
\strut
\vskip-1cm
\end{figure}
\begin{figure}[t]
\centerline{
\includegraphics[width=0.9\linewidth,clip]{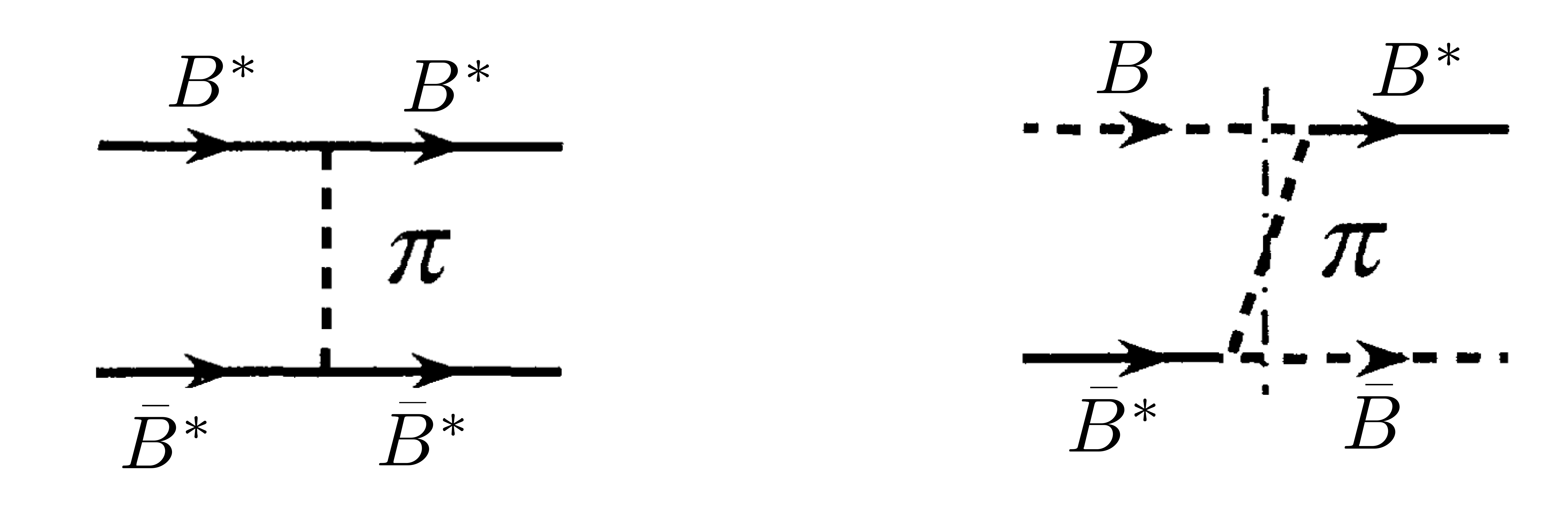}
}
\caption{Diagrams contributing to $B \bar B^*$ and $B^* \bar B^*$ binding
through pion exchange. Analogous diagrams contribute to 
$D \bar D^*$ and $D^* \bar D^*$ binding, modulo the caveat that 
$D^* \ra D\pi$, while $B^* \ra B \pi$ is kinematically forbidden,
as discussed in the text.}
\label{fig:pion_exchange}
\strut
\vskip-1cm
\end{figure}
\subsection{Interpretation as deuteron-like quasi-bound states of two
heavy mesons}
The most interesting theoretical question is what are these states?
Their quantum numbers are those of a $\bar b b \bar u d$ tetraquark, but
such quantum numbers can also be realized by a system consisting of 
$B^* \bar B$ and $B^* \bar{B^*}$ ``hadronic molecules" loosely bound
by pion exchange, as schematically shown in
Figure~\ref{fig:tetraquark_vs_molecule}.

The difference between these two
possibilities is subtle, because they have the same quantum numbers and
therefore {\em in principle} they can 
can mix with each other. The
extent of the mixing depends on the overlap between the respective
wave functions. By a ``tetraquark" I mean a state where all four quarks are
within the same ``bag" or confinement volume, while by ``hadronic molecule"
I mean a state where there are two color-singlet heavy-light mesons 
attracting each other by exchange of pions and possibly other light mesons.

The proximity of the two resonances to the 
$B^* \bar B$ and $B^* \bar{B^*}$ thresholds
strongly suggests a parallel with $X(3872)$, whose mass is almost exactly
at the $D^* \bar D$ threshold.

It also provides strong support for the the possibility that these state
indeed are 
deuteron-like ``molecules" of two heavy mesons quasi-bound by
pion exchange \cite{Voloshin:1976ap,deusons,Tornqvist:2004qy},
as schematically shown in Figure~\ref{fig:pion_exchange}. This is because
it is very unlikely that two ``genuine" tetraquarks just happened to sit at
the respective two-meson thresholds.

The attraction due to $\pi$ exchange is 3 times weaker
in the $\,I{=}1\,$ channel than in the $\,I{=}\,0$ channel.
 Consequently, in the charm system the $\,I{=}1\,$ state is 
expected to be
well above
the \,$D^* \bar D$\, threshold and the $\,I{=}0\,$ \ $X(3872)$ is
at the threshold.\footnote{For simplicity we treat $X(3872)$ 
as an isoscalar, since it
has no charged partners, and we ignore here the issue of isospin breaking in
its decays. A more refined treatment results in the same conclusions.}
In the bottom system 
the attraction due to $\,\pi\,$ exchange is essentially the same, but
the kinetic energy is much smaller
by a factor of ${\sim} m(B)/m(D){\approx} 2.8$\,.
Therefore the net binding is much stronger than in the charm system. 

The recently discovered manifestly exotic charged resonances
are surprisingly narrow. This is the case 
 in both $\bar b b$ systems
\cite{Collaboration:2011gja,Belle:2011aa}
-- and in the exotic charmonium, 
 namely the remarkable peak $Z_c(3900)$ 
at $3899.0\pm3.6\pm4.9$ MeV with $\Gamma=46\pm10\pm20$  MeV
reported by BESIII~\cite{Ablikim:2013zna}.

The relatively slow decay of these exotic resonances implies that the
dominant configuration of the $\bar Q Q \bar q q$  four-body system is
{\em not} that of a low-lying $\bar Q Q$ quarkonium and pion(s).  
The latter have a much lower energy than the respective 
two-meson thresholds $\bar M M^*$ and $\bar{M^*} M^*$, $(M=D,B$),
but do readily fall apart into $(\bar Q Q)$ and pion(s) and would result in very large 
decay widths.
Thus we should
 view these systems as loosely bound states and/or near threshold resonances in the two heavy-meson system.

Such ``molecular" states, $\bar D D^*$, etc., were introduced
in Ref.~\cite{Voloshin:1976ap}. They were  later extensively discussed 
\cite{deusons,Tornqvist:2004qy} 
in analogy with the deuteron which binds via exchange of pions and
other light mesons, and were referred to as ``{\em deusons}". 
The key observation is that the coupling to the heavy mesons of the light mesons  exchanged 
($\pi$, $\rho$, etc.) becomes universal and independent of $M_Q$ for $M_Q\to\infty$,
and so does the resulting potential in any given $J^{CP}$ and isospin channel.
In this limit the kinetic energy $\sim p^2/(M_Q)$ vanishes, 
and the two heavy mesons bind with a  binding energy $\sim$ the maximal depth of the attractive meson-exchange potential.

For a long time it was an important open question whether these 
consideration apply in the real world with large but finite masses of the 
$D$ and $B$ mesons.
The recent experimental results of Belle 
 \cite{Collaboration:2011gja,Belle:2011aa}
and  BESIII \cite{Ablikim:2013zna}, together with theoretical analysis 
in Refs.~\cite{Karliner:2011yb} and \cite{Karliner:2012pc} 
strongly indicate that 
such exotic states do exist -- some were already found and more 
are predicted below.

Due to parity conservation, the pion cannot be  exchanged  in the $\bar M M$
system,  but it does contribute in the $\bar M M^*$ and $\bar{M^*} M^*$ 
channels. This fact provides additional support to the molecular
interpretation, because resonances have been observed close to
$\bar D D^*$, $\bar D^* D^*$, $\bar B B^*$ and $\bar B^* B^*$ thresholds,
but no resonances have been seen at $\bar D D$ nor at $\bar B B$
thresholds.

The \hbox{$\vec \tau_1\cdot\vec\tau_s$} isospin nature of the exchange  implies that 
the the binding is 3 times stronger in the isoscalar channel.
It was estimated \cite{Karliner:2011yb,Karliner:2012pc}
that in the bottomonium system 
this difference in the binding potentials raises the $I{=}1$ exotics well
above the $I{=}0$ exotics. In the charmonium system this splitting 
is expected to be slightly smaller, because the $\bar D D^*$/$\bar{D^*} D^*$ 
states are larger than $\bar B B^*$/$\bar{B^*} B^*$.
This is because
the reduced mass in the $\bar B B^*$ system is approximately 2.5 times
larger than in the $\bar D D^*$ system. On the other hand, the net attractive
potential due to the light mesons exchanged between the heavy-light mesons
is approximately the same, since $m_c, m_b \gg \Lambda_{QCD}$.
As usual in quantum mechanics, for a given potential 
the radius of a bound state or a resonance gets
smaller when the reduced mass grows, so the 
$\bar D D^*$ states are larger than the $\bar B B^*$ states.
Because of this difference in size
the attraction
in both $I{=}0$ and $I{=}1$ charmonium 
channels is expected to be somewhat smaller. In addition, 
Ref.~\cite{Nussinov:1998se} suggested that the asymptotic 
coupling between the two heavy mesons and a pion,
$g_{MM^*\pi}$ for $m_Q\to\infty$ is approached from below, so that
$g_{BB^*\pi} > g_{DD^*\pi}$.

\subsection{Prediction of additional related exotic states}
Since the quarks are heavy, we can treat their kinetic energy as a
perturbation depending linearly on a parameter inversely proportional
to $\mu_{red}$, 
the reduced mass of the two meson system, which scales like the mass of
the heavy quark~\cite{Karliner:2013dqa}, with the hamiltonian
$ H= a\cdot p^2 + V$,
where 
$a={1/\mu_{red}}\sim 1/m_Q$.
We can use the existing data in order to make a very rough estimate of
the isovector binding potential in the $m_Q \to \infty$ limit.
We have two data points: $Z_c(3900)$ at
$a(D)$ is approximately 27 MeV above $\bar D D^*$ threshold and
$Z_b(10610)$ at $a(B)$ 
is approximately 3 MeV above $\bar B B^*$ threshold. Linear
extrapolation to $a =0$ yields $E_b^{I=1}(a{=}0)\approx {-}11.7$ MeV. In
view of the convexity of the binding energy in $a$, the actual binding energy is likely to slightly exceed this linear extrapolation.

We can then use this result for the isovector channel to estimate the $\bar B
B^*$ binding in the isoscalar channel. Assuming that the isoscalar binding energy
in the $m_Q\to\infty$ limit is 3 times larger than for the isovector,
we have $E_b^{I{=}0}(a{=}0) \approx 3\cdot({-}11.7)={-}35$ MeV. 
The state
\,$X(3872)$ is at $\bar D D^*$ threshold,
providing an additional data point of $E_b^{I{=}0}(a(D))\approx 0$ 
in the isoscalar
channel.
Linear extrapolation to $a(B)$ yields approximately ${-}20$
MeV as the $\bar B B^*$ binding energy in the isoscalar channel.

The upshot is that the newly discovered $Z_c(3900)$ isovector resonance
confirms and refines the estimates in
\cite{Karliner:2011yb,Karliner:2012pc} for the mass of
the putative $\bar B B^*$ isoscalar bound state. 
This immediately lead to several predictions~\cite{Karliner:2013dqa}:
\vskip-1.6em\strut
\begin{itemize}
\item[a)] two $I=0$ narrow resonances $X_b$ in the bottomonium system,
about 23 MeV below $Z_b(106010)$ and $Z_b(10650)$, i.e. about 20 MeV below 
the corresponding $\bar B B^*$ and $\bar{B^*} B^*$ thresholds;
\item[b)] an $I=1$ resonance above $\bar{D^*} D^*$ threshold;
\item[c)] an $I=0$ resonance  near $\bar{D^*} D^*$ threshold.
\end{itemize}
The $X_b$ states can most likely be observed through the decays
$X_b \ra \Upsilon \omega$ or $X_b \ra \chi_b \pi \pi$.
Unlike incorrectly stated in~\cite{Karliner:2013dqa}, they {\em cannot}
decay to $\Upsilon \pi \pi$. The latter decays are prevented by $G$-parity
conservation.\cite{Mizuk} The observed decay $X(3872) \ra J/\pi\pi$ is
only possible because isospin is strongly broken between $D^+$ and
$D^0$, and because $X(3872)$ is at the $\bar D D^*$ threshold.
In contradistinction, in the bottomonium system isospin is almost perfectly conserved.
{\em Thus the null result in CMS search \cite{Chatrchyan:2013mea} 
for $X_b \ra \Upsilon(1S) \pi^+ \pi^-$ does not tell us if $X_b$ exists.}

Quite recently the BESIII collaboration 
reported observation 
in $e^+e^- \to (D^{*} \bar{D}^{*})^{\pm} \pi^\mp$
of what looks just like (b) above, namely 
a new charmonium-like charged resonance $Z_c(4025)$, slightly
above the $\bar{D^*} D^*$ threshold, at 
$\sqrt{s}=(4026.3\pm2.6\pm3.7)$\,MeV, with width 
of $24.8\pm5.6\pm7.7$\,MeV~\cite{Ablikim:2013emm}.
Shortly afterward, \hbox{BESIII} reported observation of 
another charged charmoniumlike structure 
$Z_c(4020)$ in $e^+e^- \ra \pi^+ \pi^- h_c$
at 
$ (4022.9\pm 0.8\pm 2.7)$ MeV 
and width of $7.9\pm 2.7\pm 2.6$ MeV. At this time it is not yet
clear if these are two independent resonances, or two observations of the
same object at slightly different masses, possibly due to 
systematic effects associated with the two observation channels.

\begin{sidewaysfigure}
\centering
\includegraphics[width=1.0\linewidth,clip]{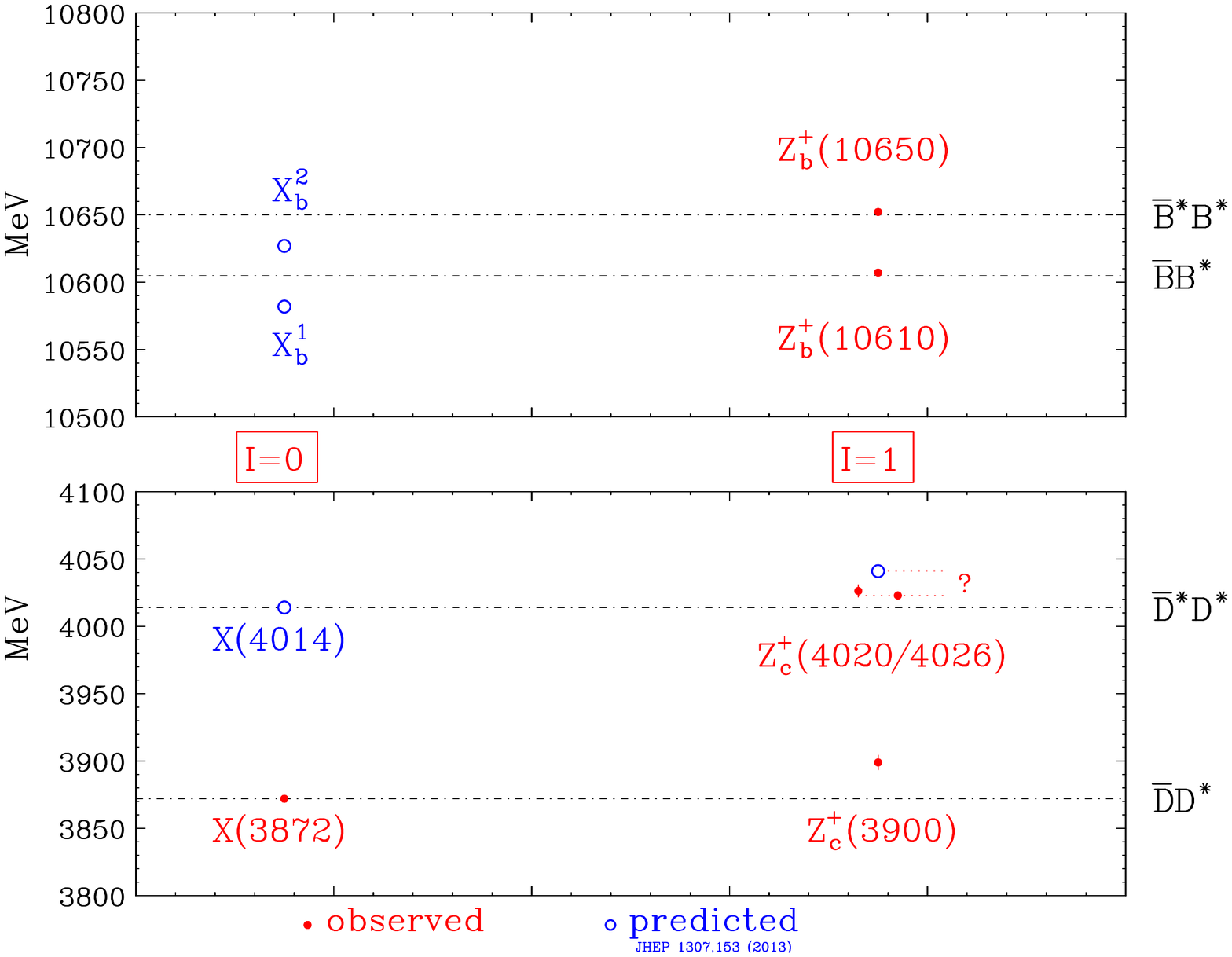}
\caption{Masses of the doubly-heavy exotic quarkonia vs. two-meson thresholds. 
The states observed so far are shown in red, the predicted states are shown in
blue. $I=0$ resonances are shown on the left and isovectors are shown on the
right. Note the proximity of all the states to the corresponding two-meson
thresholds.}
\label{fig:exotic_quarkonia}
\end{sidewaysfigure}

\begin{sidewaysfigure}
\centering
\includegraphics[width=1.0\linewidth,clip]{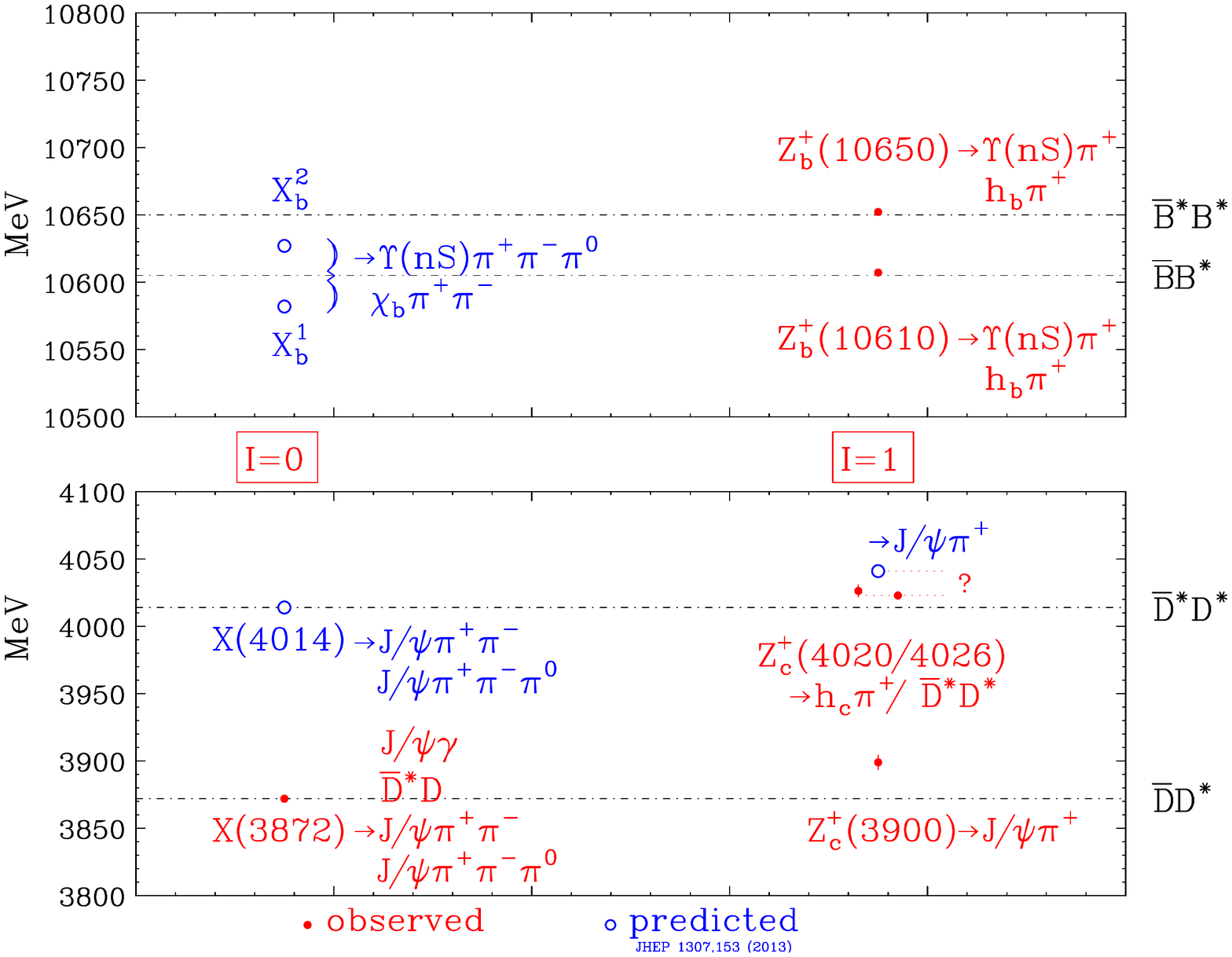}
\caption{Decay channels of doubly-heavy exotic quarkonia. The legend is as
in Figure \protect\ref{fig:exotic_quarkonia}.}
\label{fig:exotic_quarkonia_with_decay_channels}
\end{sidewaysfigure}

Figure~\ref{fig:exotic_quarkonia} provides a concise summary of the 
experimental information about the masses of doubly-heavy exotics
observed so far, together with our predictions for the masses of 
additional states, as discussed above. 
Figure~\ref{fig:exotic_quarkonia_with_decay_channels} shows the observed
and predicted decay channels of these states.

It is somewhat puzzling that, unlike $Z_c^+(3900)$,
$Z_c^+(4020/4026)$ 
has not been seen in the $J/\psi \pi^+$ mode. Moreover, one notes that 
$Z_c^+(4020/4026)$ 
is somewhat closer 
to the $\bar D^* D^*$ threshold than our prediction. It will be interesting to
identify the reasons for this small difference. The two main possibilities are 
(a) details of the experimental analysis; (b) a possible difference between the 
$\bar B^* B^*$ and $\bar D^* D^*$ attractive pion-exchange potentials.
Such a difference might perhaps be due to the fact that 
$m(B^*)- [m(B) + m(\pi)] \approx {-}94$ MeV, while
$m(D^*) - [m(D) + m(\pi)] \approx 0^\pm$, depending on the $D^*$ and $\pi$
charges, affecting energy denominators in virtual pion emission.

\subsection{{\bf \boldmath A \,$(\Sigma_b^+\Sigma_b^-)$ \,{\em beauteron}\, 
dibaryon? \unboldmath}}
The discovery of the $Z_b$-s and their interpretation as
quasi-bound \,$B^* \bar B$\, and \,$B^* \bar{B^*}$,
raises an interesting possibility of a
{\em strongly bound $\Sigma_b^+\,\Sigma_b^-$ deuteron-like state,
{\bf \em a beauteron}}~\cite{KLT}.
The $\Sigma_b$ is about 500 MeV heavier than $B^*$. 
The $\Sigma_b \Sigma_b$ 
kinetic energy is therefore significantly smaller
than that of $B \bar{B^*}$ or $B^* \bar{B^*}$.
Moreover, $\Sigma_b$ with $I{=}1$ couples more strongly
to pions than $B$ and $B^*$ with $I={1\over2}$. The opposite electric
charges of $\Sigma_b^+$ and $\Sigma_b^-$ provide 
additional 2-3 MeV of binding energy.
The heavy dibaryon bound state might be sufficiently long-lived to
be observed experimentally.
A possible decay mode of the beauteron is
$(\Sigma_b^+ \Sigma_b^-) \rightarrow \Lambda_b \,\Lambda_b \,\pi^+ \pi^-$,
which might be observable in LHCb. It should also be seen in lattice QCD.

\section{\boldmath $QQ \bar q \bar q$ \,tetraquarks\unboldmath}
The quark content of the exotic resonances observed so far is 
$\bar Q Q \bar q q$.
A very different type of exotics are the
$QQ \bar q_1 \bar q_2$ \,tetraquarks
(TQ-s)~\cite{Ader:1981db,Manohar:1992nd,Gelman,Frishman:2013qaa}.
If such states do exist, producing and discovering even the
lightest $cc\bar u \bar d$ is an extraordinary challenge.  One needs to
produce  {\em two} $\bar Q Q$  pairs 
and then rearrange them, so as to form
$QQ$ and $\bar Q \bar Q$ diquarks, rather than the more favorable
configuration of two $\bar QQ$ and color singlets.  
{\em Then}  the $QQ$ diquark needs to pick up a
$\bar u\bar d$ light diquark, rather than a $q$, to make a $QQq$ baryon,
suppressing the production rate of these TQ-s below the rate  of $QQq$
doubly-heavy baryons
production which themselves are quite hard to make, as discussed below.

A small ray of hope comes from the observation of the doubly-heavy
$B_c = (\bar b c)$ mesons \cite{BcA,Bc}, suggesting
that  simultaneous production of $\bar b b$ and $\bar c c$ pairs 
which are close to each other in space and
in rapidity and can coalesce to form doubly-heavy hadrons is not too
rare. For example CDF reported~\cite{Bc} 
$108\pm15$ candidate events in the $B_c^{\pm} \to J/\psi \pi^\pm$
channel.
This is an encouraging sign for the prospects of producing and observing the
$ccq$ and $bcs$ baryons and hopefully also the $cc\bar u\bar d$ \,TQ. 
ATLAS and CMS and especially LHCb have the best chance 
of discovering these states.

If the new TQ lies below the two heavy meson threshold, it will be stable under the 
strong interaction and will decay only weakly.\footnote{A statement slightly 
modified if the TQ mass is above the 
$2 m_D$ threshold, as it can then first decay through an exceedingly
narrow EM decay $D^*\to D^0 \gamma$
\cite{Nussinov:1998se,PDG},
followed by the weak decays of the the $D$-s \cite{Gelman}.}
A discovery of such a stable exotic hadron would be a 
spectacular finding indeed.
As we discuss below, if these tetraquarks lay above threshold, 
they still may manifest themselves as narrow $D D^*$ resonances.

The following theoretical considerations are relevant for the last
point above.
First, in clear contradistinction to the states discussed earlier, the new
TQ-s are {\em unlikely} to be molecular. This is because both heavy mesons
contain light antiquarks, rather than a $\bar q$ in one and a $q$ in the other,
causing the $\omega$ and $\rho$ exchange to be repulsive, 
rather than attractive.
With no bound  $\bar D D^*$ states, one might expect the
new $D D^*$ states to lay higher and manifest themselves 
as broad  resonances, at best. 

This bleak picture will change dramatically if the $QQ\bar u \bar d$\, TQ is {\em not}
a simple molecular state, but instead a connected color network, consisting of a
$QQ$ diquark coupled to a $\hbox{\bf 3}_c^*$, with a matching 
$\bar u \bar d$ diquark coupled to a {\bf 3}$_c$. The $QQ$ diquark will
then be bound by the large Coulombic interaction ${\cal O}(\alpha_s^2 M_Q)$, 
as discussed below.
Finding such a novel color network
would be particularly exciting \cite{Gelman},\cite{Karliner:2006hf}.  With some poetic
license, we can compare it to  the discovery of the $C_{60}$ buckyballs.

\noindent
The Coulombic binding of the $ QQ $ diquark,
\beq
E_b(QQ)=
{1\over2}\cdot{1\over2}\cdot{1\over 2}{\alpha_s^2} M_Q = {1\over8} \alpha_s^2 M_Q
\label{CoulombicBinding}
\eeq
is half as strong as that of a color-singlet $\bar Q Q$. Another
${1\over2}$ factor reflects the need to use the reduced mass, and the
${1\over2}\alpha_s^2 M_Q$  is the standard hydrogen binding formula. Thus
the heavy  diquark $QQ$ becomes infinitely deeply bound as \hbox{$M_Q \to
\infty$}. For $m_c \sim 1.6$ GeV and using 1/2 for the relevant $\alpha_s$, the
Coulombic binding $\sim 50$ MeV is  moderate.  However, the  $1\over2$ ratio of
QCD Coulomb  interaction in the diquark, versus its strength in the $\bar
Q Q $ color singlet system likely extends to {\em all} QCD interactions,
including the confining stringy potential~\cite{Nussinov:1999sx}.

If indeed the $cc\bar u\bar d$ TQ is mostly described by the new color network,
then its decay into $D D^*$ will be suppressed by the required rearrangement of
the color.

In view of the above discussion, we now focus on the question whether the
lowest state of the new color network is below or near the $D D^*$ threshold.

\subsection*{Doubly-heavy baryons}
From the point of view of QCD there is nothing exotic about baryons
containing two heavy quarks \hbox{($b$ or $c$,} generically denoted by $Q$)
and one light quark ($u$ or $d$, generically denoted by $q$).
Heavy quarks decay only by weak interaction, with a characteristic
lifetime orders of magnitude larger than the typical QCD timescale,
so from the point of view of strong interactions the $QQq$ baryons are
stable,
just like protons, neutrons and hyperons.
Thus these {\em doubly-heavy} baryons {\em must} exist.

As discussed above in the context of $QQ\bar q \bar q$ tetraquarks,
producing and discovering such states
is a significant experimental challenge.  One has to
produce  {\em two} $\bar Q Q$  pairs
which need to rearrange themselves, so as to form
$QQ$ and $\bar Q \bar Q$ diquarks, rather than the more favorable
configuration of two $\bar QQ$ color singlets.
{\em Then}  the $QQ$ diquark needs to pick up a
an light quark $q$, to make a $QQq$ baryon.

There are interesting parallels between the
$QQ\bar q\bar q$ \,TQ-s and doubly-heavy baryons $QQq$. In both types of
systems
there is a light color triplet -- a quark or an anti-diquark --
bound to a heavy diquark. Because of this similarity,
{\em experimental observation of doubly-heavy baryons is very important not
just
in its own right, but as source of extremely valuable information for
deducing the properties of the more exotic
$QQ\bar q\bar q$ tetraquarks.} Such deduction can carried out just as it
was done for
$b$-baryons.

\begin{figure}[t]
\centering
\includegraphics[width=1.0\linewidth,clip]{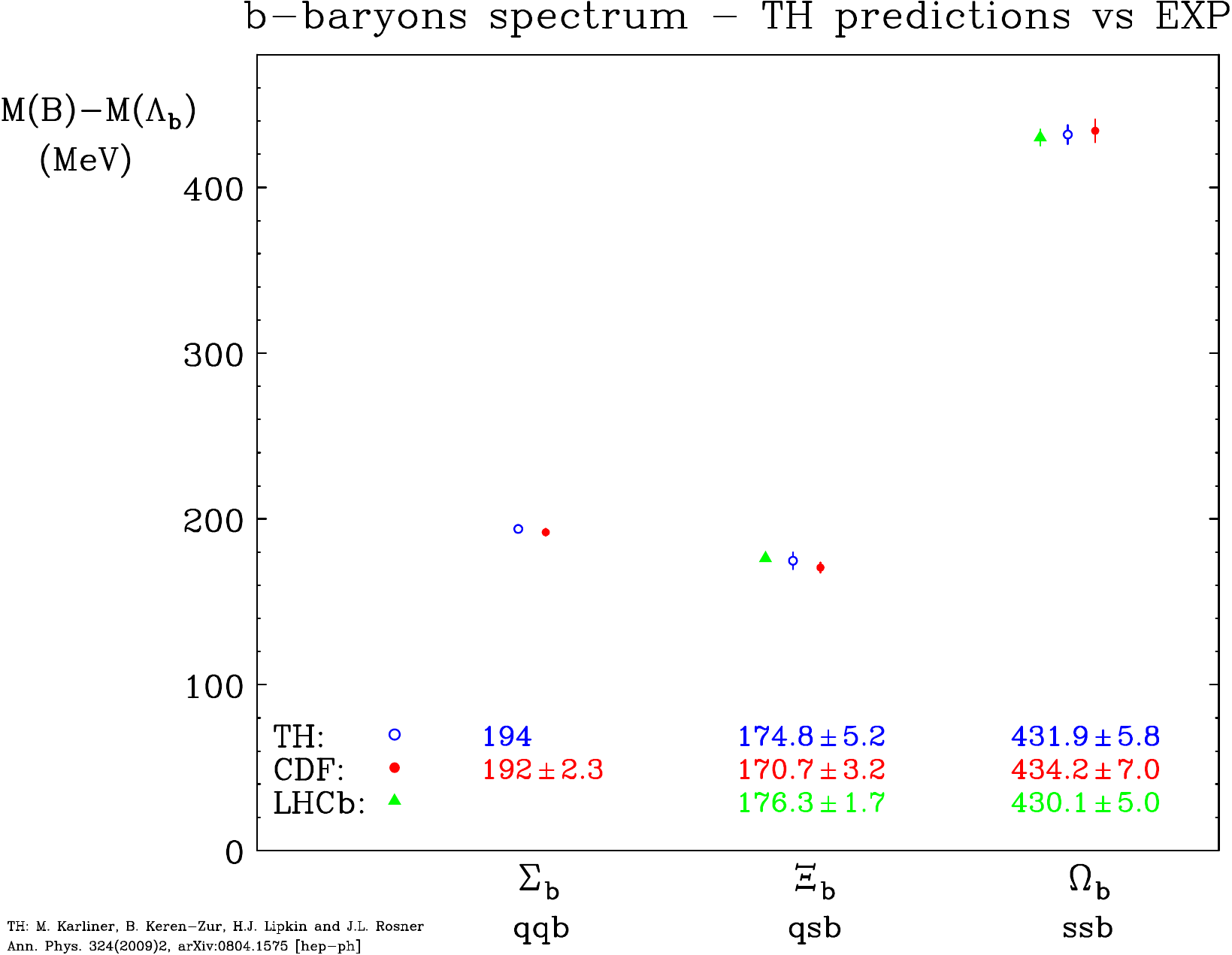}
\caption{Comparison of theoretical predictions 
\protect\cite{KLSigmab,Karliner:2007jp,Karliner:2008sv} 
and experiment for masses of $b$-baryons.}
\label{fig:bbaryons}
\end{figure}

In the last few years it became possible to accurately predict at the level 
of 2-3 MeV the masses of heavy baryons containing the $b$-quark:
$\Sigma_b(bqq)$, $\Xi_b(bsq)$ and $\Omega_b(bss)$
\cite{KLSigmab,Karliner:2007jp,Karliner:2008sv}, as shown in Figure 
\ref{fig:bbaryons}.

These predictions used as input the masses of the $B$, $B_s$, $D$ and $D_s$ mesons, together with the masses of the corresponding $c$-baryons
$\Sigma_c(cqq)$, $\Xi_c(csq)$ and $\Omega_c(css)$.

An analogous approach to the masses of the $QQ\bar q \bar q$, suggested the relation
\beq
m(c c\bar u\bar d) = m(\Xi_{ccu}) + m(\Lambda_c) - m(D^0) - {1\over4} [m(D^*)-m(D)]
\label{ccud}
\eeq
designed to optimally match interactions on both
sides~\cite{Gelman},\cite{Karliner:2013dqa}.
To date, only the SELEX experiment at Fermilab reported doubly charmed  
$ccd$ and $ccu$ baryons with  mass
 $\sim 3520$ MeV \cite{Mattson:2002vu} -- a result  not confirmed by other
experiments \cite{Aubert:2006qw,Chistov:2006zj}.  Substituting 
this mass in eq.~(\ref{ccud}) as a placeholder in a ``proof of concept" estimate
yields $M(cc\bar u\bar d)$ $\sim 3900$ MeV, which is just 30 MeV 
above the $D^* D$ threshold. 
One should keep in mind that, as implied by eq.~(\ref{CoulombicBinding}),
the binding in $(bb\bar q \bar q)$ is expected to be significantly
stronger than in $(c c\bar q\bar q)$.
In any case, reliable experimental information on $QQq$ baryon masses
is clearly essential for settling the issue.

Hadrons containing two $b$ quarks, such as double-bottom baryons $bbq$
or $b\bar b \mythinspace q \bar q$ and $b b \mythinspace \bar q \bar q$
tetraquarks have a unique and a spectacular decay mode with two
$\Jpsi$-s
in the final state. It is mediated by both $b$ quarks decaying via
$b \rightarrow \bar c c s \rightarrow \Jpsi \mythinspace s$ and yields
\beq
(bbq) \rightarrow
\Jpsi \mythinspace \Jpsi (ssq) \rightarrow \Jpsi \mythinspace \Jpsi
\mythinspace\,\Xi
\label{bbq_decay}
\eeq
\noindent
and
\beq
(b\bar b \mythinspace q \bar q) \rightarrow \Jpsi \, \Jpsi (\bar s s
\bar q q)
\rightarrow
\Jpsi \, \Jpsi \,K\, \bar K, 
\eeq
\noindent
as well as
\beq
(b b \mythinspace \bar q \bar q) \rightarrow \Jpsi \, \Jpsi (s s
\bar q \bar q)
\rightarrow
\Jpsi \, \Jpsi \,\bar K\, \bar K, 
\eeq
etc.,
with all final state hadrons coming from the same vertex. This
unique signature is however hampered by a very low rate,
expected for such a process,
especially if one uses dimuons to identify the $\Jpsi$-s.
It is both a
challenge and an opportunity for LHCb \cite{Karliner:2006hf}.

Once the double bottom baryon is identified experimentally
through (\ref{bbq_decay}) or via another channel
and its mass is measured, one will immediately be able to
estimate the mass of the corresponding 
tetraquark through the $b$-quark analogue of eq.~(\ref{ccud})
and check whether or not it is below the two $B$ meson threshold:
\begin{equation}
m(b b\bar u\bar d) = m(\Xi_{bbu}) + m(\Lambda_b) - m(B^0) - {1\over4}
[m(B^*)-m(B)]
\label{bbud}
\end{equation}

%
%
%

\end{document}